\documentclass[aps,prl,amssymb,showpacs,twocolumn]{revtex4}
\usepackage{graphicx}

\begin{document}

\preprint{submitted to PRL'08}

\title{Spontaneous Polarisation Build up in a Room Temperature Polariton Laser}

\author{J.J. Baumberg$^1$}%
\author{A.V. Kavokin$^{2}$}%
\author{S. Christopoulos$^1$}
\author{A.J.D. Grundy$^2$}%
\author{R. Butt\'{e}$^3$}%
\author{G. Christmann$^3$}%
\author{D.D. Solnyshkov$^4$}%
\author{G. Malpuech$^4$}%
\author{G. Baldassarri H$\ddot{o}$ger von H$\ddot{o}$gersthal$^2$}%
\author{E. Feltin$^3$}%
\author{J.-F. Carlin$^3$}%
\author{N. Grandjean$^3$}%
\affiliation{$^1$ NanoPhotonics Centre, Department of Physics, University of Cambridge, Cambridge, CB3 0HE}%
\affiliation{$^2$ School of Physics and Astronomy, University of Southampton, Highfield, Southampton, SO17 1BJ}%
\affiliation{$^3$ Institute of Quantum Electronics and Photonics, EPFL, 1015 Lausanne, Switzerland}%
\affiliation{$^4$ LASMEA, CNRS, 24 Avenue des Landais, 63177 Aubiere Cedex, France}%

\date{\today}

\begin{abstract}
We observe the build up of strong ($\sim$50\%) spontaneous vector polarisation in emission from a GaN-based polariton laser excited by short optical pulses at room temperature. The Stokes vector of emitted light changes its orientation randomly from one excitation pulse to another, so that the time-integrated polarisation remains zero. This behaviour is completely different to any previous laser. We interpret this observation in terms of the spontaneous symmetry breaking in a Bose-Einstein condensate of exciton-polaritons. 
\end{abstract}

\pacs{71.36.+c, 03.75.Kk, 81.07.-b, 73.21.-b}

\vspace{-0.3cm}  
\maketitle
\vspace{-0.5cm}

Polariton lasers are  coherent light sources based on emission of light from a coherent ensemble of exciton-polaritons - the mixed light-exciton quasiparticles in semiconductor microcavities. The concept of polariton lasing was first proposed in 1996 \cite{Imamoglu96}, followed a few years later by reports of coherent polariton emission in microcavities\cite{Deng02,Huynh03,Bajoni08}. Recently, we reported polariton lasing at room temperature in GaN-based microcavities\cite{Christopoulos07}. Apart from being very promising for applications, the concept of polariton lasing involves several fundamental physics issues. Contrary to conventional lasers, polariton lasers emit coherent and monochromatic light {\it spontaneously}. This is achieved when mixed light-matter quasiparticles  (exciton-polaritons), Bose-condense inside a semiconductor microcavity. Bose-Einstein condensation (BEC) of the polaritons is a subject of intense experimental and theoretical research at present. Several experimental works claiming polariton BEC have appeared recently\cite{Kasprzak06,Balili07,Lai07}. Though polariton BEC implies polariton lasing these two phenomena are not identical. For polariton lasing a macroscopically populated quantum state of exciton polaritons must be created, which can be considered as a polariton condensate. Polariton lasing does not require thermal equilibrium in the system or the spontaneous build-up of the order parameter, which are the main criteria for BEC when understood as a thermodynamic phase transition. Which experimental measurement should be considered as decisive proof for the exciton-polariton BEC is still a subject of debate within the community. Thermalisation of the exciton-polaritons detected by angle-resolved photoluminescence (PL) measurements has been considered one of the key criteria for a long time\cite{Kasprzak06,Balili07}. However, a similar angular dependence of the PL has also been observed in GaAs-based photon lasers\cite{Bajoni07}. The spatial coherence of polariton emission demonstrated in Ref. \cite{Kasprzak06} is characteristic for conventional lasers as well. Recent theoretical work suggests that observation of the spontaneous build-up of the vector polarisation in emission from polariton lasers would be evidence for the spontaneous symmetry breaking in the system\cite{Laussy06,Shelykh06,Combescot07}. In turn,  spontaneous symmetry breaking is considered to be a smoking gun for BEC ever since the pioneering work of Goldstone\cite{Goldstone61,Pitaevskii03}.

Here we report observations of the build up of the spontaneous vector polarization at room temperature in bulk GaN microcavities. Unlike the recent low temperature experiments on BEC in CdTe- and GaAs-based cavities in which the polarisation direction was pinned along the crystal axes\cite{Kasprzak06,Balili07}, we find that the polarisation of emitted light varies stochastically from one experiment to another. We also observe the threshold-like build-up of population in the lowest energy polariton state, a Boltzmann distribution of exciton-polaritons in excited states (with an effective temperature of 360K) and the build-up of first order coherence. Thus each criterion for polariton BEC formulated in previous works is fulfilled in our sample.

Intense theoretical research on Bose-Einstein condensation (BEC) from 1938-1965 led to the definition of the BEC criterion for a weakly interacting Bose gas which is associated with the appearance of a macroscopic condensate wave-function, $\psi(\underline{r})$, forming the order parameter of the phase transition. Yang\cite{Yang62} termed the phenomenon `off-diagonal long-range order'. The system Hamiltonian is invariant to the phase of $\psi(\underline{r})$, however at the phase transition the symmetric solution becomes unstable. The system therefore breaks symmetry by choosing a {\it specific} phase that is adopted throughout the whole condensate.

In this context, BEC of exciton polaritons has important specific features: the exciton-polaritons are spinor quasi-particles with two possible spin projections (up/down, corresponding to right-/left-circular polarisations of emitted light). Therefore, the order parameter for exciton-polariton BEC possesses two components: $\psi(\underline{r})= \left[
\begin{array}{l} \psi_{\uparrow}(\underline{r}) \\ \psi_{\downarrow}(\underline{r}) \\ \end{array} \right]
$, where $\psi_{\uparrow}(\underline{r})$  and $\psi_{\downarrow}(\underline{r})$  are the complex spin-up and spin-down wave functions, respectively. The 3D polarisation vector, ${\bf S}$ (termed the Stokes vector in classical optics or pseudo-spin in quantum mechanics), is linked to these wave functions through: $S_x = \Re (\psi_{\uparrow}^* \psi_{\downarrow})$, $S_y = \Im (\psi_{\uparrow}^* \psi_{\downarrow})$, and $S_z = \frac{1}{2}(|\psi_{\uparrow}|^2 - |\psi_{\downarrow}|^2)$. The absolute polarisation degree of the condensate is $\rho = |\psi^2|/N_0$ , where $N_0$ is the occupation number of the condensate. It is linked with linear, diagonal and circular polarisation degrees $\rho_l,\rho_d, \rho_c$  by $\rho = \sqrt{\rho_l^2+\rho_d^2+\rho_c^2}$, where $\rho_{l,d,c} = 2 S_{x,y,z}/N_0$. In microcavities pumped below threshold, $|\psi|=0$, while at threshold, $\psi$  builds up due to stimulated scattering of polaritons from the excited states to the condensate.\cite{Rubo04} The essential feature of a {\it bulk} microcavity is the absence of a spin quantisation axis leading to spin-isotropic polariton-polariton interactions. Extending the method of \cite{Rubo04}, the probability of realising a given value of the order parameter at a given time $P(\psi,t)$ is described by a non-linear Fokker-Planck equation
\begin{equation}
\frac{\partial P}{\partial t} = \nabla \left[ P \nabla U(\psi,t) + D(t) \nabla P \right],
\end{equation}
where the effective potential is given by
\begin{equation}
U = \left\{ \left[ W_{out}(t) - W_{in}(t) \right] | \psi |^2 + \alpha | \psi |^2 \right\}/4 ,
\end{equation}
and the diffusion coefficient $D=W_{in}(t)/4$. Here  $W_{in}(t)$ is the rate of exciton-polaritons scattering into the condensate which is dependent on the pumping strength and the coupling of the condensate to the reservoir of exciton-polaritons having large wave-vectors. $W_{out}$ is the depletion rate of the condensate depending mostly on the polariton radiative life-time, while $\alpha > 0$  is the polariton-polariton interaction constant. The population of the condensate $N_0$ is linked to $W_{in}$   and  $W_{out}$  via the Boltzmann equation
\begin{equation}
\frac{dN_0}{dt} = W_{in}(t) (N_0+1) - W_{out}(t) N_0
\end{equation}

Above threshold, $W_{in} > W_{out}$  so that the potential (2) has a maximum at $|\psi|$=0 surrounded by a circular minimum at finite $|\psi|$. This allows for an efficient diffusion of the order parameter out of the centre of the effective potential. The build-up of this order parameter explicitly results in a polarisation build-up. The resulting value of the absolute polarisation is linked to the minimum of the effective potential (2), which has no privileged polarisation, so that in each experiment the system chooses its polarisation randomly. Once formed, the polarisation continues changing slowly due to the diffusion of the order parameter within the effective potential.

\begin{figure}[!htb]
\centering
\includegraphics[width=9cm]{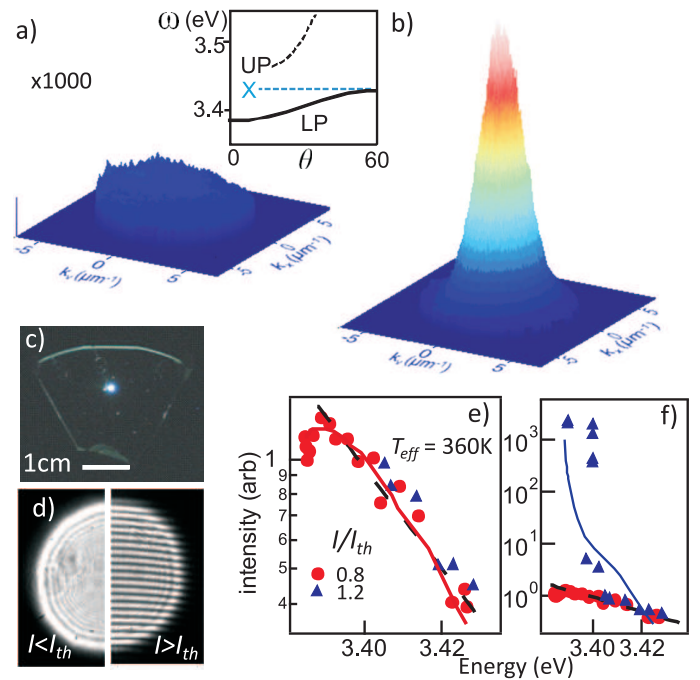}
\caption{\label{fig1} Formation of polariton lasing at $T$=300K in a GaN microcavity at in-plane wavevectors up to $k_{max}$=7 $\mu$m$^{-1}$ for (a) just below (scaled up by x1000) and (b) just above threshold ($I_{th}\sim1$mW). Inset shows dispersion. (c) Image of pumped sample above threshold. (d) Interference of far-field emission cone through a slightly-misaligned Michelson interferometer above and below threshold. (e,f) Polariton emission intensity just below and above threshold as a function of energy, together with Boltzmann fit (dashed) giving an effective temperature of 360K, and result from kinetic simulation (lines).}
\end{figure}

To demonstrate experimentally the build-up of polariton vector polarisation, we have grown a crack-free GaN-based microcavity, where the excitons are stable at 300K due to their large binding energy.
To avoid polarisation pinning which is likely to be inevitable in quantum well cavities due to their reduced symmetry, the cavity is composed of bulk GaN. In this semiconductor, the different types of excitons are mixed and the exciton-exciton interaction is expected to be spin isotropic\cite{Aoki99}  so that polariton BEC in such cavities should be accompanied by the spontaneous appearance of completely arbitrary polarisation. This device structure demonstrates strong exciton-photon coupling and polariton lasing at room temperature as we showed recently \cite{Christopoulos07}. We briefly describe the sample and experiment which is fully characterised in Ref. \cite{Christopoulos07} and references therein. Lattice-matched AlInN/AlGaN bottom and SiO$_2$/Si$_3$N$_4$ top Bragg reflectors are used to sandwich a 220nm-thick GaN active layer, forming a microcavity in the strong coupling regime displaying the classic exciton-polariton anticrossing dispersion with angle in ultraviolet emission near $\lambda$=365nm with a Rabi splitting of 36meV. The detuning varies non-monotonically across the sample and is negatively-detuned (by $<$30meV) for results reported here (but the results are not strongly detuning-dependent). The sample is pumped at an angle of 30$^{\circ}$ with 150fs non-resonant UV pulses tuned to 300nm where the top mirror becomes transparent. The experimental excitation is repeated every 4.2$\mu$s, and the emission spectra, angle-dependence and polarisation are recorded using UV-sensitive CCDs and photomultipliers (PMTs). The cone of emitted light in $k$-space (where $k$ is the in-plane wavevector) is shown in Fig.1(a,b) both below and above the threshold $I_{th}\sim 1.0$mW, showing the emergence of macroscopic occupation near $k$=0. The exponential increase of light emission from polaritons inside the microcavity at threshold manifests polariton lasing (Fig.1c).

We interfere two copies of the collimated far-field light cone in a Michelson interferometer to display the first order coherence (Fig.1d). The temporal decay of this coherence is dominated by the sub-ps emission time of the condensate. Just below threshold, the integrated polariton emission measured at each angle follows a quasi-thermal Boltzmann-like distribution with an effective temperature of $\simeq$360K$\pm$30K (Fig.1e).\cite{note19} Similar results are obtained across the wafer, with variation mainly due to the local quality factor of the Bragg mirrors.\cite{Christmann06} 

The time-integrated polarisation of emission we observe is exactly zero, independent of the orientation of the analyser axes. To resolve the polarisation of the emission for each experimental realisation of the condensation from each excitation pulse, we polarisation split the emission and focus both orthogonal components either a) onto the input slit of a Streak Camera operated in single-shot mode, or b) onto the cathodes of balanced PMTs. Previous experiments have shown the maximum lifetime of electronic excitations in the sample is 35ps (with an exponential decay time). Hence between pulses the sample completely recovers, and each pulse forms a new experiment. Light emerging up to  $\pm$15$^{\circ}$ to the sample normal is collected using a UV achromatic lens and collimated. The light passes through broadband UV waveplates (half-wave and quarter-wave) before being split by polarisation beamsplitters (care has to be taken over the angular acceptance and alignment of these optics). In the experiments which use repeated measurements of one particular polarisation basis, a single polarisation beamsplitter is used, and the two emerging beams are individually focussed on two detectors. For Streak camera measurements, they are focussed onto different positions on the input slit of the Streak camera, and their optical paths made equal so that the two polarisation components of each pulse appear at the same time position on the Streak image. The Streak camera is manually triggered, with a gain sufficient to record the relatively weak emission, allowing extraction of each polarisation component averaged over each laser shot. Equivalently, we also use UV-sensitive photomultipliers instead of the Streak camera, with a temporal response of $<$100ns, and after pre-amplification and pulse clean up electronics, these are recorded using a PC-based ADC card with sample-and-hold facility which records simultaneous measurements on all channels. The overall time resolution of the system is $<$1$\mu$s, and allows extraction of the intensity of each polarisation component for each pulse, over thousands of successive shots (Fig.2). We always use the wave plates to swap the polarisation basis between PMTs allowing calibration of the relative gain of each channel. By splitting the emission into two extra beams using two additional polarisation-independent beamsplitters at near normal incidence, we are able to simultaneously record 4 PMTs (for instance along Horizontal, Vertical, Diagonal, Right-circular bases) allowing complete reconstruction of the full Stokes parameters. We also use linearly polarised input light to calibrate the system and confirm that no artefacts are present from the polarisation splitting optics. In particular, we confirm that reversing the polarisation axes (for instance swapping $H$ and $V$) on the detectors, produces exactly equivalent results.

\begin{figure}[t]
\centering
\includegraphics[width=9cm]{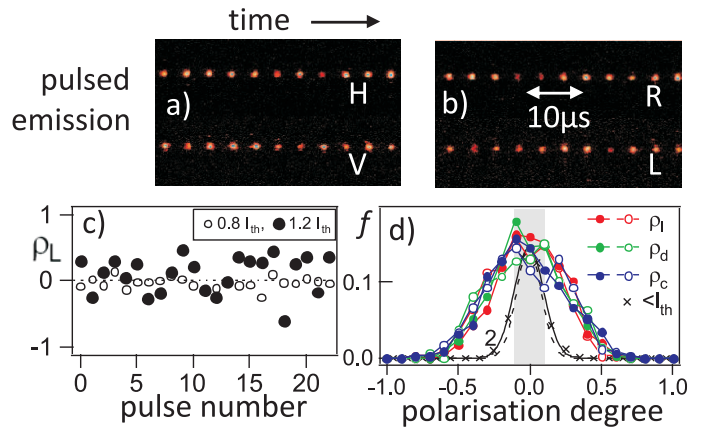}
\caption{\label{fig2} Above-threshold polarisation-resolved emission when analysing along (a) Horiz/Vert and (b) Right/Left bases. (c) Extracted linear polarisation degree showing stochastic variation from pulse to pulse above and below threshold. (d) Histogram of the fraction of each polarisation state, $f$, along linear, diagonal and circular bases of nearly 2000 polariton condensates. Open and closed circles show repeated measurement with reversed polarisation split (e.g $H$/$V$ and $V$/$H$), while crosses show below-threshold unpolarised emission statistics ($\div$2, within detection sensitivity shaded grey and scaled dashed line).}
\end{figure}

Electronic noise in the detection produces an apparent background polarisation magnitude in addition to that contained within the optical signal. Hence even a perfectly polarised input produces a polarisation histogram with some deviation about $\rho_l$=0. For the signal to noise in our measurements this is  $\sim$8\% and is shown shaded grey in Fig.2(d) and Fig.3(d).

Using appropriate waveplates we can thus examine the polarisation of each pulse along linear ($H$,$V$), diagonal ($D$,$\overline{D}$) and circular ($R$,$L$) bases. The fidelity of these measurements is calibrated using the linearly-polarised pump laser for different orientations of the input $\lambda$/2 plate, and exceeds 95\%. Comparable data from the above-threshold polariton laser [Fig.2(a,b)] reveals that each pulse of BEC has a {\it different} polarisation. The intensity of each polarisation component is extracted and used to calculate the linear polarisation degree (Fig.2c), $\rho_l = (I_H - I_V)/(I_H + I_V)$, with equivalent measures for diagonal ($\rho_d$) and circular ($\rho_c$) bases. Repeating the experiment for several hundred pulses in each polarisation basis allows construction of a polarisation histogram (Fig.2d). Below threshold, the microcavity emission is unpolarised exhibiting thermal noise around zero absolute polarisation. Above threshold, the emission is found to be instantaneously polarised, but with no preferential orientation. The magnitude of the mean polarisation is 25\% for each basis (given by the standard deviations in Fig.2d), giving a total mean polarisation of  $\sim$43\% just above threshold. The mean polarisation measured for a single shot is $<$100\% due to the random walk of the polarisation vector on a time scale given by the coherence time of the condensate (Fig.3a). 

\begin{figure}[t]
\centering
\includegraphics[width=9cm]{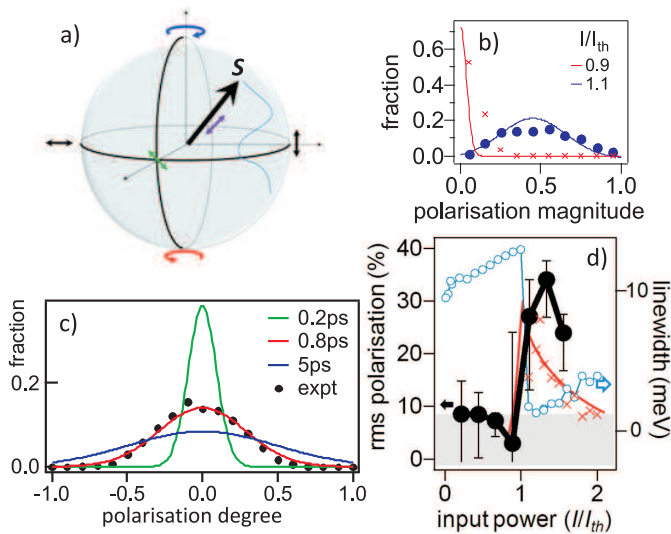}
\caption{\label{fig3}(a) Diffusion of the pseudospin, ${\bf S}$ (Stokes polarisation vector) of a polariton condensate. (b) Experimental (points) and modelled (lines) histogram of polarisation vector lengths, showing shift from zero to $\sim$ 43\% above threshold. (c) Statistical distribution of the linear polarisation degree of the condensate showing expt vs theory for different dephasing times. (d) Root mean square linear polarisation degree as a function of pump power (black). Result of theory described in the text (crosses, red line is guide to eye) uses linewidth (blue) as input parameter.}
\end{figure}

In order to reproduce the experimental polarization histogram, we perform a Monte Carlo simulation of the microcavity emission. At each moment of time the condensate is assumed to be completely polarized but the direction of its pseudospin ${\bf S}$ (see Fig.3a) is initially random. The polarisation has a probability to change its direction which is inversely proportional to the dephasing time of the condensate $t_i$. The coherence time is given by\cite{Laussy06} $t_c = \left( \frac{W_{in}}{N_0+1} + \frac{1}{t_i} \right)^{-1}$. $W_{in}$ and  $N_0$ are time dependent and are calculated using the semi-classical Boltzmann equations (3) describing polariton relaxation (see \cite{Solnyshkov07}). The dephasing time $t_i$  is governed by the fluctuation of the number of polaritons in the system. For cw pumping $t_i \simeq \frac{\hbar}{V \sqrt{N}}$  where $V$ is the matrix element of interparticle interactions and $N$  the average number of particles. To simulate the experiment with pulsed excitation, the polarized emission of the ground state is averaged for 100 ps after each pulse and the numerical experiment is repeated 10$^5$ times.

This model is analogous to the classical Heisenberg spin model\cite{Joyce67} used to describe the magnetization in ferromagnets. The spontaneous build-up of polarisation is clearly observed above threshold (Fig.3b). The dephasing time used in the calculation is directly taken from the linewidth measurements (Fig.3d). Examining the polarisation histograms (Fig.3c) shows the best fit is obtained for $t_i$ = 0.8ps (indeed corresponding to the resolution-limited 0.82 meV linewidth). Below threshold, the residual polarisation is below the detection limit of our setup, while it increases to a maximum at threshold (Fig.3d) before decreasing again because of the shortening of the dephasing time, due to stronger polariton-polariton interactions.\cite{Porras03}

These results thus show that the bosonic exciton-polaritons in GaN which form a phase-coherent state above a characteristic density, exhibit spontaneous symmetry breaking above 300K. This contrasts to linear polarisation seen in all previous systems (including both InP-based\cite{Sceats99} and similar GaN-based\cite{Chu06} bulk VCSELs), and is directly observed through the polarisation of light emitted by polaritons. The coherent polariton state thus fulfils all the criteria to be classed as a Bose-Einstein condensate. We stress that these results are completely different from all previous observations in lasers.

This work was supported by EU STIMSCAT, EPSRC Portfolio, the Swiss National Science
Foundation, and the Sandoz Family Foundation.

\vspace{-0.5cm}




\begin{references}
\bibitem{note19}  We restrict this analysis to angles below 30$^{\circ}$ because at higher angles the top Bragg mirror shifts nearly out of resonance. In this pulsed regime, the time-integrated emission intensity at each angle is directly proportional to the polariton occupation at the equivalent in-plane $k$.
\end{references}
\end{document}